\documentclass[final]{ias2}

\usepackage{graphicx}
\usepackage{multirow}
\usepackage{array}
\usepackage{hyperref}

\newcommand{\be}{\begin{equation}}
\newcommand{\ee}{\end{equation}}
\newcommand{\bea}{\begin{eqnarray}}
\newcommand{\eea}{\end{eqnarray}}
\newcommand{\ba}{\begin{array}}
\newcommand{\ea}{\end{array}}
\newcommand{\xbf}[1]{\mbox{\boldmath $#1$}}

\begin{document}

\markboth{Quadrupole moments of low lying baryons}{Neetika Sharma and Harleen Dahiya}

\title{Quadrupole moments of low lying baryons with spin $\frac{1}{2}^+$,
spin $\frac{3}{2}^+$, and spin $\frac{3}{2}^+ \to \frac{1}{2}^+$
transitions}
\author[sin]{Neetika Sharma}
\author[sin]{Harleen Dahiya}
\email{dahiyah@nitj.ac.in}
\address[sin]{Department of Physics, Dr. B.R. Ambedkar National
Institute of Technology, Jalandhar, 144011, India}
\begin{abstract}
The chiral constituent quark model ($\chi$CQM) with general
parameterization (GP) method has been formulated to calculate the quadrupole moments of the spin
$\frac{3}{2}^+$ decuplet baryons and spin $\frac{3}{2}^+ \to \frac{1}{2}^+$
transitions. The implications of such a model have been
investigated in detail for the effects of symmetry breaking and GP
parameters pertaining to the two- and three-quark contributions. It
is found that the $\chi$CQM is successful in giving a quantitative
and qualitative description of the quadrupole moments.
\end{abstract}



\maketitle

\section{Introduction}
One of the main challenge in the theoretical and experimental
hadronic physics is to understand the  structure of hadrons within the Quantum Chromodynamics (QCD) \cite{sach}, which can be obtained from the measurements of electromagnetic form factors. Following the discoveries that the quarks and antiquarks carry only 30\% of the total proton spin \cite{emc}, the orbital angular momentum of quarks and gluons is expected to make significant contribution. In addition to this, there is a significant contribution coming from the strange quarks in the nucleon which are otherwise not present in the valence structure. It therefore becomes interesting to discuss the interplay between the spin of non valence quark and the orbital angular momentum in understanding the spin structure of baryons. Further, the experimental developments \cite{kelly,glashow}, providing
information on the radial variation of the charge and magnetization
densities of the proton, give the evidence for the deviation of the
charge distribution from spherical symmetry. On the other hand it is well known that the quadrupole moment of the nucleon should vanish on account of its
spin-1/2 nature. This observation has naturally turned to be the
subject of intense theoretical and experimental activity.

The $\Delta$(1232) resonance is the lowest-lying excited
state of the nucleon in which the search for quadrupole strength has been carried out \cite{blanpied,tia03}. The spin and parity selection rules in the $\gamma + p \to \Delta^+$
transition allow three contributing photon absorption amplitudes, the magnetic dipole
$G_{M1}$, the electric quadrupole moment $G_{E2}$, and the charge
quadrupole moment $G_{C2}$. The
$G_{M1}$ amplitude gives us information on magnetic moment whereas
the information on the intrinsic quadrupole moment can be obtained
from the measurements of $G_{E2}$ and $G_{C2}$ amplitudes \cite{pdg}. If the charge distribution of the initial and final
three-quark states were spherically symmetric, the $G_{E2}$ and
$G_{C2}$ amplitudes of the multipole expansion would be zero \cite{becchi}. However, the recent experiments at Mainz, Bates, Bonn,
and JLab Collaborations \cite{jlab2,ber03} reveal that these
quadrupole amplitudes are clearly non-zero \cite{pdg}. The
ratio of electric quadrupole amplitude to the magnetic dipole
amplitude is at least $\frac{E2}{M1} \equiv - 0.025 \pm 0.005$ and a
comparable value of same sign and magnitude has been measured for
the $\frac{C2}{M1}$ ratio \cite{pdg}. Further, the quadrupole
transition moment ($Q^{\Delta^+ N}$) measured by LEGS \cite{blanpied} and Mainz \cite{tia03}
Collaborations ($-0.108 \pm 0.009 \pm 0.034 ~{\rm fm}^2$  and $-0.0846 \pm 0.0033 ~{\rm fm}^2$
respectively) also leads to the conclusion that the nucleon and the
$\Delta^+$ are {\it intrinsically} deformed.

There have been a lot of theoretical investigations in understanding
the implications of the $\frac{C2}{M1}$ and $\frac{E2}{M1}$ ratios
in finding out the exact sign of deformation in the spin $\frac{1}{2}^+$ octet baryons.
However, there is little consensus between the results even with
respect to the sign of the nucleon deformation. Some of
the models predict the deformation in nucleon as oblate \cite{oblate}, some predict a prolate nucleon deformation \cite{prolate,buch2002} whereas others speak about ``deformation''
without specifying the sign. It is important to mention here that the
deformation of the octet baryons is an intrinsic phenomena and has to be
coupled with the orbital angular momentum to compare with the spectroscopic
value. The electromagnetic
structure of the $\frac{3}{2}^+$ decuplet baryons have also been studied using the
variants of the constituent quark model (CQM) \cite{leonard,kri}, chiral quark soliton model
($\chi$QSM) \cite{soliton-model}, spectator quark model \cite{spectator}, slow rotator approach (SRA) \cite{schwesinger}, skyrme model \cite{abada}, general
parametrization method \cite{gpm}, light cone QCD sum rules (QCDSR) \cite{lcqcdsr}, large $N_c$ \cite{largenc}, chiral perturbation
theory ($\chi$PT) \cite{butler} and lattice QCD (LQCD) \cite{lein},
etc.. In this case also, the results for different theoretical
models are not consistent in terms of sign and magnitude with each
other.

One of the important nonperturbative approach in the low energy
regime of QCD is the chiral constituent quark model
($\chi$CQM) \cite{manohar}. The $\chi$CQM coupled
with the ``quark sea'' generation through the chiral fluctuation of
a constituent quark into a GBs, successfully explains the ``proton
spin crisis'' \cite{hd}, hyperon $\beta$ decay parameters \cite{nsweak},
strangeness content in the nucleon \cite{hdasymmetry}, and in the $N$ and $\Delta$ magnetic moments \cite{hds}. Further, the $\chi$CQM  coupled with the ``quark sea''
polarization and orbit angular momentum of the quark sea (referred as
the Cheng-Li Mechanism) it is able to give a
excellent fit to the $\frac{1}{2}^+$  octet and $\frac{3}{2}^+$  decuplet baryon magnetic moments and
a perfect fit to the violation of Coleman-Glashow sum rule \cite{hdmagnetic}. The model has been
successfully extended to explain the
the magnetic moment of charmed baryons with spin
$\frac{1}{2}^+$, spin $\frac{3}{2}^+$, spin $\frac{3}{2}^+ \to
\frac{1}{2}^+$ and spin $\frac{1}{2}^+ \to
\frac{1}{2}^+$  radiative decays \cite{nsmagnetic}.  In view
of the above developments, it becomes desirable to
calculate the quadrupole moment of the $\frac{1}{2}^+$  octet and $\frac{3}{2}^+$  decuplet
baryons within framework of $\chi$CQM.

The purpose of the present communication is to present a unified
approach to calculate the quadrupole moment of the spin $\frac{3}{2}^+$ decuplet baryons
including the spin $\frac{3}{2}^+ \to \frac{1}{2}^+$ transitions
within the framework of $\chi$CQM using a general parametrization
(GP) method \cite{morp89}. In particular, we aim to predict the sign of
deformation in the baryons as prolate or oblate.

\section{Quadrupole moment}
The charge radii ($r^2_B$) and quadrupole moments ($Q^B$) are the
lowest order moments of the charge density $\rho$ in a low-momentum
expansion. The charge radii contain fundamental information about
the possible ``size'' of the baryons whereas the ``shape'' of a
spatially extended particle is determined by its quadrupole moment \cite{buch01}. With respect to the body frame
of axis  the {\it intrinsic} quadrupole moment is defined as \be  \label{intrinsicqm}
Q_0 = \int d^3 {\xbf r}
\rho({\xbf r})(3z^2 - r^2)\,. \ee For the charge density
concentrated along the $z$-direction, the term proportional to
$3z^2$ dominates, $Q_0$ is positive, and the particle is prolate
shaped. If the charge density is concentrated in the equatorial
plane perpendicular to $z$ axis, the term proportional to $r^2$ dominates, $Q_0$ is negative and the particle is oblate shaped.

The {\it intrinsic} quadrupole moment $Q_0$ must be distinguished
from the {\it spectroscopic} quadrupole moment $Q$ measured in
the laboratory frame. A simple example will illustrate
this point. Suppose one has determined the quadr\-upole moment $Q_0$
of a classical charge distribution $\rho(\bf{r})$ with symmetry axis $z$
and angular momentum $J$ in the body-fixed frame according
to Eq. (\ref{intrinsicqm}).
The quadrupole moment of the same
charge distribution with respect to the laboratory frame
is then given by
\be
\label{classical}
Q= P_2(\cos \theta)\, Q_0   = \frac{1}{2} \left
(3 \, \cos^2(\theta)-1 \right ) \, Q_0
=  \left ( \frac{3 {J_{z'}}^2  - J(J+1)}{2 J (J+1)} \right )
\, Q_0,
\ee
where $\theta$ is the angle between the body-fixed $z$ and the
laboratory frame $z'$ axes, and $P_2$ is the second Legendre polynomial.
The latter arises when transforming the spherical harmonic
of rank 2 in $Q_0$ from body-fixed to laboratory coordinates.
The third equality in Eq. (\ref{classical}) is obtained
when $\cos(\theta)$ is expressed in terms of the spin projection $J_{z'}$
on the laboratory frame $z'$-axis and the total spin $J$ of the system
as $\cos(\theta)=J_{z'}/\sqrt{J^2}$. Thus, in the laboratory, one does not measure the intrinsic
quadrupole moment directly but only its projection onto the $z'$ axis.
In the quantum mechanical analogue of Eq. (\ref{classical})
the denominator of the projection factor
is changed into $(2J +3)(J+1)$.
The projection factor shows that $J=0$ and $J=1/2$ systems
have vanishing spectroscopic quadrupole moments even though
they may be deformed and their intrinsic quadrupole moments are
nonzero. Some information on the shape of nucleon or any other member of the octet baryon can be obtained by electromagnetically exciting the the baryon to the spin $J=3/2$ or higher spin state.

The most general form of the quadrupole operator composed of a two-
and three- quark term in spin-flavor space can be expressed as \cite{morp89} \bea \widehat{Q} & = & {\mathrm B'} \sum_{i \ne j}^3
e_i \left( 3 \sigma_{i \, z} \sigma_{ j \, z} - \xbf{\sigma}_i \cdot
\xbf{\sigma}_j \right) + {\mathrm C'} \!\! \sum_{i \ne j \ne k }^3
e_i \left( 3 \sigma_{j \, z} \sigma_{ k \, z} - \xbf{\sigma}_j \cdot
\xbf{\sigma}_k \right) \,, \label{quad} \eea where $\sigma_{i \, z}$ the is the $z$ component of the spin operator of the quark $i$ and $e_i$ is the quark charge. The GP method constants
${\mathrm B'}$ and ${\mathrm C'}$ parameterize the orbital and color matrix elements.
The order of  GP parameters  ${\mathrm B}$, and
${\mathrm C}$, corresponding to the two- and three-quark
terms, decreases with the increasing complexity of terms and obey
the hierarchy $|\mathrm B|> |{\mathrm C}|$.
These are fitted by using the available experimental values for the
charge radii and quadrupole moment of nucleon as input.

The quadrupole moment operators for the spin $\frac{1}{2}^+$, spin
$\frac{3}{2}^+$ baryons, and spin $\frac{3}{2}^+ \to \frac{1}{2}^+$
transitions can be calculated from the operator in Eq. (\ref{quad})
and are expressed as \bea \widehat{Q}^{B} &=& {\mathrm B'} \left( 3
\sum_{i \neq j} e_i \xbf{\sigma_{iz}} \xbf{\sigma_{jz}} - 3 \sum_{i}
e_i \xbf{\sigma_{iz}} + 3 \sum_{i} e_i \right)  \nonumber \\&& +{\mathrm C'} \left(
3\sum_{i \ne j \ne k} e_i \xbf{\sigma_{jz}} \xbf{\sigma_{kz}} + 3
\sum_{i} e_i \xbf{\sigma_{iz}} \right), \label{q1/2} \\
\widehat{{Q}}^{B^*} &=& {\mathrm B'} \left( 3\sum_{i \neq j} e_i
\xbf{\sigma_{iz}} \xbf{\sigma_{jz}} - 5 \sum_{i} e_i
\xbf{\sigma_{iz}} + 3 \sum_{i} e_i \right) \nonumber \\&& +{\mathrm C'} \left(
3\sum_{i \ne j \ne k} e_i \xbf{\sigma_{jz}} \xbf{\sigma_{kz}} + 5
\sum_{i} e_i \xbf{\sigma_{iz}} -6 \sum_{i} e_i \right),
\label{q3/2} \\ \widehat{Q}^{B^* B} &=& 3 {\mathrm B'}\sum_{i
\neq j} e_i \xbf{\sigma_{iz}} \xbf{\sigma_{jz}} + 3{\mathrm C'}
\sum_{i \ne j \ne k } e_i \xbf{\sigma_{jz}} \xbf{\sigma_{kz}}.
\label{q3/21/2} \eea It is clear from the above equations that the
determination of quadrupole moment basically reduces to the
evaluation of the flavor $(\sum_i e_{i})$, spin $(\sum_i e_i
\xbf{\sigma_{iz}})$ and tensor terms $(\sum_i e_i \xbf{\sigma_{iz}}\xbf{\sigma_{jz}})$
and $(\sum_i e_i \xbf{\sigma_{jz}} \xbf{\sigma_{kz}})$ for a given baryon.

Using the three-quark spin-flavor wave functions for the spin
$\frac{1}{2}^+$ octet and spin $\frac{3}{2}^+$ decuplet baryons, the
quadr\-upole moment can now be calculated by evaluating the matrix
elements of operators in Eqs. (\ref{q1/2}), (\ref{q3/2}) and
(\ref{q3/21/2}). We now have \bea Q^{B} &=& \langle B |\widehat{Q}_B
| B \rangle \,,\nonumber \\ Q^{B^*} &=& \langle {B^*} |
\widehat{Q}{_{B^*}} | B^* \rangle \,, \nonumber \\ Q^{B \to B^*} &=&
\langle {B^*} | \widehat{Q}{_{B^*B}}| B \rangle \,, \label{expect2}
\eea where $|B \rangle$ and $|B^* \rangle$ respectively, denote a
spin-flavor baryon wavefunction for the spin $\frac{1}{2}^+$ octet
and spin $\frac{3}{2}^+$ decuplet baryons \cite{yaouanc}.

\section{Naive Quark Model (NQM)}
The appropriate operators for the spin and flavor structure of
baryons in NQM \cite{nqm} are defined as \be\sum_i e_i = \sum_{q=u,d,s}
n^B_{q}q + \sum_{{\bar q} = {\bar u}, {\bar d}, {\bar s}}
n^B_{{\bar q}} {\bar q} = n^B_{u}u + n^B_{d}d + n^B_{s}s +
n^B_{\bar u}{\bar u} + n^B_{ \bar d}{\bar d} + n^B_{\bar s}{\bar
s} \label{numei} \,, \ee and  \be \sum_i {e_i}{\sigma_{iz}} =
\sum_{q=u,d,s} (n^B_{q_{+}}q_{+} + n^B_{q_{-}}q_{-}) =
n^B_{u_{+}}u_{+} + n^B_{u_{-}}u_{-} + n^B_{d_{+}}d_{+} +
n^B_{d_{-}}d_{-} + n^B_{s_{+}}s_{+} + n^B_{s_{-}}s_{-}
\label{numeisi} \,, \ee where $n_q^B$ $(n_{\bar q}^B)$ is the
number of quarks with charge $q$ (${\bar q}$) and $n^B_{q_{+}}$
($n^B_{q_{-}}$) is the number of polarized quarks $q_{+}$
($q_{-}$). For a given baryon $u = -\bar u$ and $u_{+} = -u_{-}$,
with similar relations for the $d$ and $s$ quarks.

For the spin $\frac{1}{2}^+$ octet baryons, the quadrupole moment of
$p$ and $\Sigma^+$ in NQM can be expressed as \bea {Q}^p
&=& 3{\mathrm B'} \left( 2u + d - 2u_+ - d_+ \right) + {\mathrm C'}
\left( -4u + d + 4u_+ - d_+ \right) \,, \label{nqmo1} \\
{Q}^{\Sigma^+} &=& 3{\mathrm B'} \left( 2u + s - 2u_+ - s_+ \right)
+ {\mathrm C'} \left(-4u + s + 4u_+ - s_+ \right) \,. \label{nqmo2} \eea

For the spin $\frac{3}{2}^+$ decuplet baryons, the quadrupole moment of $\Delta^+$ and
$\Xi^{*-}$ can be expressed as \bea {Q}^{{\Delta^+}} &=& {\mathrm
B'}\left( 6u + 3 d + 2u_+ + d_+ \right) + {\mathrm C'}
\left( -6 u - 3d + 10 u_+ + 5 d_+ \right) \,, \label{nqmd1}\\
{Q}^{{\Xi^{*-}}} &=& {\mathrm B'}\left( 3d + 6 s + d_+ + 2 s_+
\right) + {\mathrm C'}\left( -3d - 6s + 5 d_+ + 10 s_+ \right)
 \,. \label{nqmd2}\eea

Similarly, for the spin $\frac{3}{2}^+ \to \frac{1}{2}^+$ transitions, the quadrupole moment of the ${\Delta^+ p}$ and
${\Sigma^{*-} \Sigma^{-}}$
transitions can be expressed as \bea {Q}^{\Delta^+ p} &=& {2
\sqrt{2}}{\mathrm B'} \left( u_+ - d_+ \right) + {2
\sqrt{2}}{\mathrm C'} \left(-u + d \right) \,, \label{nqmt1}
\\ Q^{\Sigma^{*-} \Sigma^{-}} &=& {2 \sqrt 2} {\mathrm B'}  \left(
d_+ - s_+ \right) + {2 \sqrt 2} {\mathrm C'}\left( - d + s \right) \,.
\label{nqmt2} \eea The expressions for quadrupole moment of the
other $\frac{1}{2}^+$  octet and $\frac{3}{2}^+$  decuplet baryons, and for spin $\frac{3}{2}^+ \to
\frac{1}{2}^+$
transitions in NQM can similarly be calculated. The results are presented in
Tables \ref{cqmo}, \ref{cqmd}, and \ref{cqmt}.

\section{Chiral constituent quark model ($\chi$CQM)}
In light of the recent developments and successes of the $\chi$CQM
in explaining the low energy phenomenology \cite{hd,nsweak,hdasymmetry,hds,hdmagnetic}, we formulate the quadrupole
moments for the $\frac{3}{2}^+$  decuplet baryons and spin $\frac{3}{2}^+ \to
\frac{1}{2}^+$
transitions. The basic process in the $\chi$CQM \cite{manohar} is
the Goldstone Boson (GB) emission by a constituent quark which
further splits into a $q \bar q$ pair as \be q_{\pm} \rightarrow
{\rm GB}^{0} + q^{'}_{\mp} \rightarrow (q \bar q^{'}) +
q_{\mp}^{'}\,, \label{basic} \ee where $q \bar q^{'} + q^{'}$
constitute the ``quark sea'' \cite{cheng}.

The effective Lagrangian describing the
interaction between quarks and a nonet of GBs is \be {\cal L} = g_8
\bar q \Phi q \,, \ee with
\bea
 \Phi = \left( \ba{ccc} \frac{\pi^0}{\sqrt 2}
+\beta\frac{\eta}{\sqrt 6}+\zeta\frac{\eta^{'}}{\sqrt 3} & \pi^+
  & \alpha K^+   \\
\pi^- & -\frac{\pi^0}{\sqrt 2} +\beta \frac{\eta}{\sqrt 6}
+\zeta\frac{\eta^{'}}{\sqrt 3}  &  \alpha K^0  \\
 \alpha K^-  &  \alpha \bar{K}^0  &  -\beta \frac{2\eta}{\sqrt 6}
 +\zeta\frac{\eta^{'}}{\sqrt 3} \ea \right)~~ {\rm and} ~~q
=\left( \ba{c} u \\ d \\ s \ea \right)\,. \eea where $\zeta=g_1/g_8$, $g_1$ and $g_8$ are the coupling constants for the
singlet and octet GBs, respectively.
If the parameter
$a(=|g_8|^2$) denotes the transition probability of chiral
fluctuation of the splitting $u(d) \rightarrow d(u) + \pi^{+(-)}$,
then $\alpha^2 a$, $\beta^2 a$ and $\zeta^2 a$ respectively, denote
the probabilities of transitions of~ $u(d) \rightarrow s  +
K^{-(o)}$, $u(d,s) \rightarrow u(d,s) + \eta$, and $u(d,s)
\rightarrow u(d,s) + \eta^{'}$. SU(3) symmetry breaking is
introduced by considering $M_s > M_{u,d}$ as well as by considering
the masses of GBs to be nondegenerate $(M_{K,\eta} > M_{\pi}$ and
$M_{\eta^{'}} > M_{K,\eta})$ \cite{hd,cheng,vento}.

A redistribution of the flavor and spin structure takes place in the
interior of baryons due to the chiral symmetry breaking and the
modified flavor and spin structure can be calculated by substituting
for every constituent quark \bea q & \to & P_q q + |\psi(q)|^2\,,
\label{flavor}\\ q_{\pm} & \to & P_q q_{\pm} + |\psi(q_{\pm})|^2 \,,
\label{spin} \eea where $P_q$ is the transition
probability of no emission of GB from any of the $q$ quark and
$|\psi(q)|^2$ ($|\psi(q_{\pm})|^2$) is the transition probability of
the $q$ ($q_{\pm}$) quark, which have been detailed in Ref. \cite{hd}.

After the inclusion of ``quark sea'', the quadrupole moment for the
spin $\frac{1}{2}^+$ octet baryons vanishes on account of
the effective cancelation of contribution coming
from the ``quark sea'' and the orbital angular momentum as observed spectroscopically.
For the spin $\frac{3}{2}^+$ decuplet baryons, the quadrupole moment in $\chi$CQM can be obtained by
substituting Eqs. (\ref{flavor}) and (\ref{spin}) for each quark in
Eqs. (\ref{nqmo1}) and (\ref{nqmo2}). The quadrupole moment
of ${\Delta^+}$ and $\Xi^{*-}$ in $\chi$CQM can be expressed
as \bea Q^{\Delta^{+}} = 4{\mathrm B'} + 2{\mathrm C'} - ({\mathrm
B'} + 5 {\mathrm C'}) a\left(2 + \frac{\beta^2}{3} + \frac{2
\zeta^2}{3} \right)\,, \eea \be \ Q^{\Xi^{*-}}  = - 4{\mathrm
B'}- 2 {\mathrm C'} +  ({\mathrm B'} + 5 {\mathrm C'}) a \left(
\frac{4 \alpha^2}{3} + \beta^2 + \frac{2 \zeta^2}{3} \right)\,.\ee

Similarly, the quadrupole moment of ${\Delta^+ p}$ and
${\Sigma^{*-} \Sigma^{-}}$ transitions in $\chi$CQM can be expressed as
\bea {Q}^{\Delta^{+} p} & =& 2 {\sqrt 2} \left[ {\mathrm B'} \left(1
- a \left(1 + \alpha^2 + \frac{\beta^2}{3} + \frac{2
\zeta^2}{3} \right) \right)-{\mathrm C'} \right] \,, \\
{Q}^{\Sigma^{*-} \Sigma^{-}} & = & 2 {\sqrt 2}  {\mathrm B'}
 a \left (\frac{\alpha^2}{3} - \frac{\beta^2}{3}  \right) \,. \eea The expressions for
the quadrupole moment of other $\frac{3}{2}^+$ decuplet baryons and
spin $\frac{3}{2}^+ \to
\frac{1}{2}^+$ transitions in $\chi$CQM can similarly be calculated. The results are presented in Tables \ref{cqmd} and
\ref{cqmt}.

\section{Results and Discussion}
\label{result quad} The quadrupole moment calculations involve two set of parameters, the GP method parameters ($\mathrm B'$ and $\mathrm C'$) and the SU(3) symmetry breaking parameters of $\chi$CQM ($a$, $a \alpha^2$, $a \beta^2$, and $a \zeta^2$). In order to obtain a simultaneous fit to charge radii and quadrupole moments, the GP parameters were fitted to the presently available data for the charge radii of nucleon and quadrupole moment of $\Delta^+ N$ transition as input. The set of parameters obtained after $\chi^2$ minimization are as
follows \be {\mathrm B'} =-0.047 \,, ~~~~ {\mathrm C'} = -0.008
\,,\ee obeying the hierarchy $|{\mathrm B'}| > |{\mathrm C'}|$ \cite{morp99} corresponding to the two- and three- quark contribution. For the $\chi$CQM parameters, a best fit is obtained
by carrying out a fine grained analysis of the spin and flavor
distribution functions of proton \cite{hd,nsweak,hdasymmetry,hds,hdmagnetic}
leading to \be a = 0.12 \,, ~~ \alpha = 0.7 \,, ~~ \beta = 0.4 \,,
~~ \zeta = -0.15 \,.\ee It would be interesting to mention here that in the present work
we have used the same set of $\chi$CQM parameters used to calculate the magnetic moment of baryons \cite{hd,hdmagnetic}.

Using the set of parameters discussed above, we have calculated the
numerical values for the quadrupole moment for the $\frac{3}{2}^+$  decuplet
baryons in $\chi$CQM and presented the results in Table \ref{quado}.
The results of the spin $\frac{3}{2}^+ \to
\frac{1}{2}^+$ transitions have been presented
in Table \ref{quadt}. To understand the implications of chiral symmetry
breaking and ``quark sea'', we have also presented the results of
NQM. Since the calculations in $\chi$CQM have been carried out using
the GP method, the NQM results have also been presented by including
the two-, and three- quark term contributions of the GP parameters
so that the contribution of the ``quark sea'' effects can be
calculated explicitly. For the case of spin $\frac{1}{2}^+$ octet baryons, we find that
the quadrupole moments are zero for all the cases in NQM. Even if we
consider the contribution coming from two-quark terms with the inclusion of
``quark sea'', and SU(3) symmetry breaking effects the
quadrupole moments still remain zero. Because of the angular momentum conservation, the quadrupole moments of particles with total angular momentum 1/2 are zero and remain zero, no matter how complicated the operator or wave function. However, the non zero value can be obtained for the `intrinsic' quadrupole moments. Further, the inclusion of orbital angular momentum contribution would lead to a zero spectroscopic quadrupole moment because the ''quark sea'' spin and orbital angular momentum contribute with the opposite sign. This also agrees with the experimental observations.
In the case
of $\frac{3}{2}^+$  decuplet baryons, the quadrupole moments of the charged baryons
are equal whereas all neutral baryons have zero quadrupole moment.

For the $\frac{3}{2}^+$  decuplet, and radiative decays of baryons, it can be
easily shown that in the SU(3) symmetric limit, the magnitude of quadrupole moments can be
expressed by the following relations \bea
\frac{Q^{\Delta^{++}}}{2} &=& Q^{\Delta^{+}}= Q^{\Delta^{-}} =
Q^{\Sigma^{*+}} = Q^{\Sigma^{*-}} = Q^{\Xi^{*-}} =
Q^{\Omega^-}\,,\eea
\be Q^{\Delta^{+} p} = Q^{\Sigma^{*+} \Sigma^{+}}  = 2 Q^{\Sigma^{*0}
\Sigma^{0}}  = Q^{\Xi^{*0} \Xi^{0}} = \frac{2}{\sqrt 3}
Q^{\Sigma^{*0} \Lambda}\,. \ee The inclusion of SU(3) symmetry
breaking changes this pattern considerably and we obtain \bea
Q^{\Omega^-} > Q^{\Xi^{*-}} > Q^{\Sigma^{*-}} >
Q^{\Delta^-} > \frac{Q^{\Delta^{++}}}{2} > Q^{\Delta^{+}} >
Q^{\Sigma^{*+}} \,, \nonumber \eea \be  2 Q^{\Sigma^{*0} \Sigma^{0}} >
Q^{\Xi^{*0} \Xi^{0}} = Q^{\Sigma^{*+} \Sigma^{+}}  > Q^{\Delta^{+}
p} = \frac{2}{\sqrt 3} Q^{\Sigma^{*0} \Lambda} \,.\ee Also we have
\be Q^{\Xi^{*0}} =  Q^{\Sigma^{*0}} = Q^{\Delta^0}, \ee which has
its importance in the isospin limit where the three-quark core in
neutral baryons does not contribute to the quadrupole moment. In the
limit of SU(3) symmetry breaking, a non vanishing value for the
neutral baryons quadrupole moment is generated by the ``quark sea''
through the chiral fluctuations of constituent quarks leading to \be
Q^{\Xi^{*0}} > Q^{\Sigma^{*0}} > Q^{\Delta^0}. \ee In the
SU(3) limit, the transition moments involving the negatively charged
baryons are zero \be  Q^{\Xi^{*-} \Xi^{-}} = Q^{\Sigma^{*-}
\Sigma^{-}} = 0.\ee This is because, if flavor symmetry is exact, U-spin
conservation forbids such transitions.
The exact order of SU(3) symmetry breaking effects can be easily
found from Tables \ref{quado} and \ref{quadt}. Since there is no
experimental or phenomenological information available for any of
these quadrupole moments, the accuracy of these relations can be
tested by the future experiments.

For the spin $\frac{3}{2}^+$ decuplet baryons presented in Table
\ref{quado}, quadrupole moments results in NQM using the GP method
predict an oblate shape for all positively charged baryons
($\Delta^{++}$, $\Delta^+$, and $\Sigma^{*+}$), prolate shape for
negatively charged baryons ($\Delta^-$, $\Sigma^{*-}$, $\Xi^{*-}$,
and $\Omega^-$). It is important to mention here that the NQM is
unable to explain the deformation in neutral baryons ($\Delta^0$,
$\Sigma^{*0}$, and $\Xi^{*0}$). On incorporating the effects of
chiral symmetry breaking and ``quark sea'' in the $\chi$CQM, a small
amount of prolate deformation in neutral baryons ($\Delta^0$,
$\Sigma^{*0}$, and $\Xi^{*0}$) is observed. The trend of
deformations is however the same for the positively and negatively
charged baryons in $\chi$CQM and NQM. The other phenomenological
models also observe a similar trend, for example, light cone QCD sum
rules \cite{lcqcdsr}, spectator quark model \cite{spectator},
lattice QCD \cite{lein}, $\chi$PT \cite{butler}, chiral quark
soliton model ($\chi$QSM) \cite{soliton-model}, etc..

For the case of spin $\frac{3}{2}^+ \to
\frac{1}{2}^+$ transitions in Table \ref{quadt},
it is observed that  quadrupole moments of all the transitions are
oblate in shape. This result is further endorsed by the predictions
of Skyrme model \cite{abada}. The effects of chiral symmetry
breaking can further be substantiated by a measurement of the other
transition quadrupole moments.

\section{Summary and Conclusion}
To summarize, $\chi$CQM is able to provide a fairly good description
of the quadrupole moments of spin
$\frac{3}{2}^+$ decuplet baryons and spin $\frac{3}{2}^+ \to
\frac{1}{2}^+$ transitions using general parameterization (GP)
method. For the spin $\frac{3}{2}^+$ decuplet baryons prolate shape is observed for the neutral decuplet baryons
($\Delta^0$, $\Sigma^{*0}$, and $\Xi^{*0}$). This trend of
deformation is same for the positively and negatively
charged baryons in $\chi$CQM as well as other phenomenological
models, for example, light cone QCD sum
rules \cite{lcqcdsr}, spectator quark model \cite{spectator},
lattice QCD \cite{lein}, $\chi$PT \cite{butler}, chiral quark
soliton model ($\chi$QSM) \cite{soliton-model}, etc.. For the case of spin $\frac{3}{2}^+ \to
\frac{1}{2}^+$ transitions, the oblate shape of the quadrupole moments of all the transitions are
endorsed by the predictions
of Skyrme model \cite{abada}. New experiments aimed
at measuring the quadrupole moments of other baryons as well as transitions would provide
us with valuable information on the structure of the baryons which
is needed for a profound understanding of the nonperturbative regime
of QCD.

\section*{Acknowledgments}
This work was partly supported by Department of Science and Technology, Government of India (SR/S2/HEP-0028/2008) and Department of Atomic Energy, Government of India (2010/37P/48/BRNS/1445).

\begin{table}[ph]
{\footnotesize \caption{Quadrupole moments of the octet baryons in NQM using GP method.}
{\label{cqmo}
\begin{tabular}{@{}cc@{}}
\hline
Baryons & NQM \\ \hline

$p$ & $ 3{\mathrm B'} \left(2 u + d - 2 u_{+} - d_{+} \right) +
{\mathrm C'} \left(-4 u + d + 4 u_{+} - d_{+} \right)$ \\

$n$ & $ 3{\mathrm B'} \left( u + 2 d - u_{+} - 2d_{+} \right) +
{\mathrm C'} \left( u - 4 d - u_{+} + 4 d_{+} \right)$ \\

$\Sigma^+$ & $ 3{\mathrm B'} \left(2 u + s - 2 u_{+} - s_{+} \right)
+ {\mathrm C'} \left(-4  u + s + 4 u_{+} - s_{+} \right)$ \\

$\Sigma^-$ & $3{\mathrm B'} \left( 2 d + s - 2 d_{+} - s_{+} \right)
+ {\mathrm C'} \left(-4  d + s + 4 d_{+} - s_{+} \right)$ \\

$\Sigma^0 $ & $3{\mathrm B'} \left(u + d + s - u_{+} - d_{+} - s_{+}
\right) $ \\&
 $+ {\mathrm C'} \left(- 2 u - 2 d + s + 2 u_{+} + 2 d_{+} -
s_{+} \right)$ \\

$\Xi^0 $ & $3{\mathrm B'} \left(u + 2s - u_{+} - 2s_{+} \right) +
{\mathrm C'} \left(u - 4s - u_{+} + 4 s_{+} \right)$ \\

$\Xi^-$ & $3{\mathrm B'} \left( d + 2s - d_{+} - 2s_{+} \right)
+{\mathrm C'} \left(d -4s - d_{+} + 4 s_{+} \right)$ \\

$\Lambda^0 $ & $3{\mathrm B'} \left(u + d + s - u_{+} - d_{+} -
s_{+} \right) +3 {\mathrm C'} \left(-s + s_{+} \right)$\\
\hline
\end{tabular}
}}

\end{table}

\begin{table}[ph]
{\footnotesize \caption{Quadrupole moments of the decuplet baryons in NQM and $\chi$CQM
using GP method.}
{\label{cqmd}
\begin{tabular}{@{}ccc@{}}
\hline
Baryon & NQM & $\chi$CQM \\ \hline

$\Delta^{++}$ & $ {\mathrm B'} \left(9u +3 u_{+} \right) + {\mathrm
C'} \left(- 9 u + 15 u_{+} \right)$ & $ 8{\mathrm B'} + 4{\mathrm
C'} - ({\mathrm B'} + 5 {\mathrm C'}) \frac{a}{3}
\left(9 + 3 \alpha^2 + 2 \beta^2 + 4 \zeta^2 \right) $\\

$\Delta^{+}$ & ${\mathrm B'} \left(3(2 u +d)+ 2 u_{+} + d_{+}
\right) + {\mathrm C'} \left(-3(2 u+ d) + 5( 2u_{+} + d_{+})
\right)$ & $4{\mathrm B'} + 2{\mathrm C'} - ({\mathrm B'} + 5
{\mathrm C'}) \frac{a}{3} \left(6 + \beta^2 + 2 \zeta^2 \right)$\\

$\Delta^{0}$ & ${\mathrm B'} \left( 3(u+ 2 d) + u_{+} + 2 d_{+}
\right) + {\mathrm C'} \left(-3(u+ 2d) + 5(u_{+} +2 d_{+}) \right)$
& $({\mathrm B'} + 5 {\mathrm C'})a \left(-1 + \alpha^2 \right)$ \\

$\Delta^{-}$ & ${\mathrm B'} \left( 9d + 3d_{+} \right) + {\mathrm
C'} \left(- 9 d + 15 d_{+} \right)$ & $-4{\mathrm B'} - 2 {\mathrm
C'} + ({\mathrm B'} + 5 {\mathrm C'})\frac{a}{3} \left(6 \alpha^2 +
\beta^2 + 2 \zeta^2 \right)$ \\

$\Sigma^{*+}$ & ${\mathrm B'} \left( 3(2 u + s) + 2 u_{+} + s_{+}
\right) + {\mathrm C'} \left(-3(2 u+ s) + 5( 2u_{+} + s_{+})
\right)$ & $ 4{\mathrm B'} + 2{\mathrm C'} - ({\mathrm B'} +
5{\mathrm C'}) \frac{a}{3} \left( 6 + \alpha^2 + 2 \zeta^2 \right) $\\

$\Sigma^{*-}$ & ${\mathrm B'} \left( 3(2 d + s) + 2 d_{+} + s_{+}
\right) + {\mathrm C'} \left(-3(2 d + s) + 5(2 d_{+} + s_{+})
\right)$ & $-4{\mathrm B'} - 2{\mathrm C'} + ({\mathrm B'} + 5
{\mathrm C'}) \frac{a}{3} \left(5 \alpha^2 + 2 \beta^2 + 2 \zeta^2 \right)$ \\

$\Sigma^{*0}$ & ${\mathrm B'} \left( 3(u + d + s) + u_{+} + d_{+} +
s_{+} \right)$ & $({\mathrm B'} + 5 {\mathrm C'})\frac{a}{3} \left(
-3 + 2 \alpha^2 + \beta^2 \right)$\\
& $+ {\mathrm C} \left(-3(u + d + s) + 5(u_{+} + d_{+} +
s_{+}) \right)$&\\

$\Xi^{*0}$ & ${\mathrm B'} \left(3(u + 2 s) + u_{+} + 2s_{+} \right
) + {\mathrm C'} \left(-3(u + 2s) + 5(u_{+} + 2s_{+}) \right)$ &
$({\mathrm B'} + 5 {\mathrm C'}) \frac{a}{3} \left( -3 + \alpha^2 +
2 \beta^2 \right) $\\

$\Xi^{*-}$ & ${\mathrm B'} \left( 3(d + 2 s) + d_{+} + 2s_{+}
\right) + {\mathrm C'} \left(-3(d + 2s) + 5(d_{+} + 2s_{+}) \right)$
& $- 4{\mathrm B'} - 2 {\mathrm C'} + ({\mathrm B'} + 5 {\mathrm
C'})\frac{a}{3} \left( 4 \alpha^2 + 3 \beta^2 + 2 \zeta^2 \right)$\\

$\Omega^{-}$ & ${ \mathrm B'}\left( 9s + 3 s_{+} \right) + {\mathrm
C'} \left(- 9 s + 15 s_{+} \right)$ & $-4{\mathrm B'} - 2 {\mathrm
C'} + ({\mathrm B'} + 5 {\mathrm C'}) \frac{a}{3} \left(3\alpha^2 +
4 \beta^2 + 2\zeta^2 \right)$ \\ \hline
\end{tabular}
}}

\end{table}

\begin{table}[ph]
{\footnotesize \caption{Quadrupole moments of the spin $\frac{3}{2}^+ \to \frac{1}{2}^+$
transitions in NQM and $\chi$CQM using GP method.}
{\label{cqmt}
\begin{tabular}{@{}ccc@{}}
\hline
Baryon &  NQM & $\chi$CQM \\ \hline
${\Delta^{+} p}$  & ${2 \sqrt 2} {\mathrm B'} \left(u_{+} - d_{+}
\right) +{2 \sqrt 2} {\mathrm C'}\left(-u + d \right)$ & $ 2{\sqrt
2} {\mathrm B'} \left(1 - \frac{a}{3}(3 + 3 \alpha^2 + \beta^2 + 2
\zeta^2) \right) -2 \sqrt{2} {\mathrm C'} $ \\

${\Sigma^{*+} \Sigma^{+}}$ & ${2 \sqrt 2} {\mathrm B'}  \left( u_{+}
-s_{+} \right) +{2 \sqrt 2} {\mathrm C'}\left(-u + s \right)$ & $ 2
{\sqrt 2} {\mathrm B'} \left(1 - \frac{a}{3}(3 + 2 \alpha^2 + 2
\beta^2 + 2 \zeta^2) \right) -2 {\sqrt 2} {\mathrm C'}$ \\

${\Sigma^{*-} \Sigma^{-}}$  & ${2 \sqrt 2} {\mathrm B'} \left(d_{+}
-s_{+} \right) + {2 \sqrt 2} {\mathrm C'}  \left( -d + s \right)$&
$\frac{2 {\sqrt 2}}{3}{\mathrm B'} a \left( \alpha^2 - \beta^2
\right)$\\

${\Sigma^{*0} \Sigma^{0}}$ & ${ \sqrt 2} {\mathrm B'} \left( u_{+} +
d_{+} -2s_{+} \right) +{ \sqrt 2} {\mathrm C'} \left( -u - d + 2 s
\right) $ & $ {\sqrt 2} {\mathrm B'} \left(1 - a (1 +
\frac{\alpha^2}{3} + \beta^2 + \frac{2}{3} \zeta^2)
\right) - {\sqrt 2} {\mathrm C'}$ \\

${\Xi^{*0} \Xi^{0}}$ & ${2 \sqrt 2} {\mathrm B'} \left(u_{+} - s_{+}
\right) + {2 \sqrt 2} {\mathrm C'} \left( - u + s \right)$ & $ 2
{\sqrt 2} {\mathrm B'} \left(1 - \frac{a}{3} (3 + 2 \alpha^2 + 2
\beta^2 + 2 \zeta^2) \right) - 2 {\sqrt 2} {\mathrm C'}$ \\

${\Xi^{*-} \Xi^{-}}$ & ${2 \sqrt 2} {\mathrm B'} \left(d_{+} - s_{+}
\right) + {2 \sqrt 2} {\mathrm C'}  \left(-d + s \right)$ &
$\frac{2{\sqrt 2}}{3}  {\mathrm B'} a \left( \alpha^2 - \beta^2
\right)$ \\

$\Sigma^{*0} \Lambda$ & ${\sqrt 6}{\mathrm B'} \left(u_{+} - d_{+}
\right) +{\sqrt 6} {\mathrm C'} \left(-u + d \right)$ & $
\sqrt{6}{\mathrm B'} \left( 1 - \frac{a}{3} (3 + 3 \alpha^2 +
\beta^2 + 2 \zeta^2) \right) -\sqrt{6} {\mathrm C'} $ \\ \hline
\end{tabular}
}}

\end{table}
\begin{table}[ph]
\caption{Quadrupole moments of the spin $\frac{3}{2}^+$  decuplet
baryons in $\chi$CQM using GP method
and SU(3) symmetry breaking .}
{\begin{tabular}
{@{}ccc ccc ccc@{}} \hline
Baryon & NQM &CQM &$\chi$PT&SRA& Skyrme& GPM&
\multicolumn{2}{c}{$\chi$CQM with SU(3)} \\ \cline{8-9}
& & \cite{kri} & \cite{butler}& \cite{schwesinger}& \cite{abada} &
\cite{gpm}& symmetry& symmetry\\
&&& &&& && breaking\\
&  fm$^2$ & $10^{-2}$
fm$^2$&$10^{-1}$
fm$^2$& $10^{-1}$
fm$^2$&  $10^{-2}$
fm$^2$&fm$^2$&fm$^2$&fm$^2$ \\\hline
${\Delta^{++}}$ & $-$0.409 &$-$9.3&$-$0.8 $\pm$ 0.5&$-$0.87&$-$8.8
& $-$0.12 & $-$0.3437 & $-$0.3695 \\
${\Delta^{+}}$& $-$0.204 &$-$4.6&$-$0.3$\pm$0.2 &$-$0.31&$-$2.9
& $-$0.06 & $-$0.1719 & $-$0.1820 \\
${\Delta^{0}}$ &0.0 &0.0&0.12 $\pm$0.05&0.24&2.9
&0.0&0.0& 0.0055 \\
${\Delta^{-}}$ & 0.204 &4.6&0.6 $\pm$0.3&0.80&8.8
&0.06&0.1719& 0.1930 \\
${\Sigma^{*+}}$ & $-$0.204 &$-$5.4&$-$0.7 $\pm$0.3&$-$0.42&$-$7.1
& $-$0.069& $-$0.1719& $-$0.1808 \\
${\Sigma^{*-}}$ & 0.204 &4.0& 4.0 $\pm$ 0.2&0.52&7.1
&0.039&0.1719& 0.1942 \\
${\Sigma^{*0}}$ & 0.0 &$-$0.7&$-$0.13 $\pm$0.07&0.05&0.0
&0.014&0.0& 0.0067 \\
${\Xi^{*0}}$ &0.0 &$-$1.3&$-$0.35 $\pm$0.2&$-$0.07&$-$4.6
&$-$0.1719&0.0& 0.0079 \\
${\Xi^{*-}}$ & 0.204 &3.4&0.2 $\pm$0.1&0.35&4.6
&0.024&0.1719& 0.1954 \\
${\Omega^-}$ & 0.204 &2.8& 0.09$\pm$0.05&0.24&0.0
&0.014&0.1719& 0.1966 \\\hline
\end{tabular}
\label{quado}
}
\end{table}

\begin{table}[ph]
\caption{Quadrupole moments of the spin $\frac{3}{2}^+ \to
\frac{1}{2}^+$ decuplet to octet transitions in $\chi$CQM using GP
method and SU(3) symmetry breaking.}
{\label{quadt}
\begin{tabular}{@{}ccc ccc@{}}
\hline
Baryon & NQM &Skyrme& GPM& \multicolumn{2}{c}{$\chi$CQM with SU(3)} \\
\cline{5-6}

& & \cite{abada}& \cite{gpm}& symmetry & symmerty\\
&& && &breaking\\
  &fm$^2$ &$10^{-2}$ fm$^2$& fm$^2$& fm$^2$& fm$^2$ \\\hline
${\Delta^{+} p}$ & $-$0.110&$-$5.2& $-$0.082& $-$0.0608
& $-$0.0846 \\
$-0.0846 \pm 0.0033$&& && & \\

${\Sigma^{*+} \Sigma^{+}}$ & $-$0.110 &$-$0.93& $-$0.076& $-$0.0608&
$-$0.0864 \\
${\Sigma^{*-} \Sigma^{-}}$ & 0.0 &0.93& 0.014& $-$0.0608& $-$0.0018 \\
${\Sigma^{*0} \Sigma^{0}}$ & $-$0.055 &0.0& $-$0.031& 0.0& $-$0.0441 \\
${\Xi^{*0} \Xi^{0}}$ & $-$0.110 &2.91& $-$0.031&$-$0.0608& $-$0.0864 \\
${\Xi^{*-} \Xi^{-}}$ & 0.0 &$-$2.91& 0.007&0.0& $-$0.0018 \\
${\Sigma^{*0} \Lambda} $ & $-$0.096&$-$4.83&$-$0.041& $-$0.0526& $-$0.0733
\\
\hline

\end{tabular}}
\end{table}

\end{document}